\documentclass[%
 reprint,
superscriptaddress,
 amsmath,amssymb,
 aps,
pre,
longbibliography
]{revtex4-2}

\usepackage{graphicx}% Include figure files
\usepackage{dcolumn}% Align table columns on decimal point
\usepackage{bm}% bold math
\usepackage[normalem]{ulem} % RS: I changed the style to keep the italics

\usepackage{cancel}  % Include this in the preamble

\usepackage{xcolor}

\definecolor{darkgreen}{rgb}{0,0.4,0.0}

\begin{document}

\title{ Parameter degeneracy in the vertex model for tissues}

\author{Paulo C. Godolphim}
\affiliation{Departamento de F\'\i sica, FCFM, Universidad de Chile, Santiago, Chile}
\affiliation{Instituto de Física, Universidade Federal do Rio Grande do Sul, Porto Alegre, Brasil.}

\author{Leonardo G. Brunnet}
\affiliation{Instituto de Física, Universidade Federal do Rio Grande do Sul, Porto Alegre, Brasil.}

\author{Rodrigo Soto}%
\affiliation{Departamento de F\'\i sica, FCFM, Universidad de Chile, Santiago, Chile}

\date{\today}% It is always \today, today,

\begin{abstract}
The vertex model with homogeneous cell properties is known to exhibit a parameter degeneracy in which the system’s dynamics is independent of the target area. Here, we show, for the heterogeneous vertex model where cells differ in size and stiffness, that degeneracy is also present with the average product of target areas and stiffness becoming dynamically irrelevant. 
Fixing this quantity is equivalent to fixing the global internal tissue pressure. Unless properly treated, this degeneracy undermines the physical relevance of key observables’ numerical values, such as cell target shape index, cell pressure, and cell stress tensor. We present methods to resolve the degeneracy and to correctly set the gauge pressure via symmetry transformations applied to the cells’ target areas. We further demonstrate that the degeneracy is removed under certain boundary conditions and partially lifted when spherical tissues are modeled using a locally planar approximation, leading to 
numerical consequences when fitting model parameters to experimental data. The approach extends beyond vertex models and provides a framework for testing whether the parameter spaces of other physical models are free from degeneracy.
\end{abstract}

\maketitle

\section{Introduction}
The physics of tissues is fundamental for understanding several biological processes, such as morphogenesis, wound healing, embryonic development, and the origin of diseases \cite{Heisenberg2013,Goodwin2021,Di2023}. To mechanically describe these systems, a variety of models have been proposed \cite{Camley2017}, including approaches such as the Potts model \cite{Kabla2012}, deformable particle models \cite{ Teixeira2021, Ourique2022, Teixeira2024}, and phase field models \cite{Chiang2024}. Among these alternatives, cell edge network models \cite{Dow2023} remain among the most widely used, standing out due to their simplicity and ability to represent the geometry and topology of cells in tissues. 
Moreover, with the advancement of quantitative biology \cite{Dow2023} and the goal of obtaining reliable quantitative results from models and experimental data, cell edge network models serve as the physical basis for most of the force inference techniques \cite{Ishihara2012,Chiou2012,Vanslambrouckid2024,Jurado2025,Roffay2021}. These methods allow for the extraction of the mechanical properties of tissues directly from experimental images.

Among the cell edge network models, the first to be widely adopted was the vertex model (VM) \cite{Fletcher2014,Alt2017}, originally proposed as an energy functional in Ref.~\cite{Nagai2001} and popularized in Ref.~\cite{Farhadifar2007}.  
Since then, various extensions have emerged, such as the self-propelled Voronoi (SVP) model~\cite{Bi2016,Yang2017}, in which the degrees of freedom lie in the cell centers rather than in the vertices, the active tension network model~\cite{Noll2017}, which incorporates sub-rules of interaction in the cellular edges, and more recent formulations, such as the active foam model \cite{Kim2021}, which allows the existence of curved membranes and extracellular space between cells, or the introduction of plasticity and activity to cells~\cite{Barton2017,Verdugo2022}.
The most common form of the energy functional for the VM is
\begin{equation}\label{eq-vertexEnergyHomogeneous}
    E^\text{homo} = \sum_c \left [ \frac{K_{A}}{2}\left (A_c - A_{0} \right )^2 + \frac{K_{L}}{2}(L_c - L_{0})^2 \right],
\end{equation}
where the sum is over all cells, $A_c$ and $L_c$ are cell areas and perimeters, $A_{0}$ and $L_{0}$ are target areas and perimeters, and $K_{A}$ and $K_{L}$ are the area and perimeter stiffnesses. Here, the vertices evolve to minimize $E^\text{homo}$, driving the cells to approach their areas and perimeters to the target values, when possible. Equation~\eqref{eq-vertexEnergyHomogeneous} defines the homogeneous model, where all cells share the same parameters, i.e., they have identical physical properties. The VM has been shown to capture relevant qualitative behaviors of epithelial tissues, such as rigidity transitions and cell rearrangements~\cite{Farhadifar2007, Staple2010}, and has been used to describe several biological systems, including cone photoreceptor organization in Zebrafish \cite{Salbreux2012}, \textit{Fundulus} epiboly \cite{Weliky1990}, active pulses in Killifish epiboly \cite{Verdugo2022}, notochord development in \textit{Xenopus} \cite{Weliky1991}, polarity in mouse blastocyst \cite{Honda2008}, wing development in \textit{Drosophila}~\cite{Mao2013,Sui2018,Tetley2019}, and more~\cite{Fletcher2014, Alt2017}.

For VMs, studies show that in certain regions of parameter space (known as floppy states), the energy minimum corresponds to a continuous manifold in configuration space rather than a single point ~\cite{Farhadifar2007,Hoevar2009,Noll2017}.
This allows for configuration changes, such as cell intercalation \cite{Farhadifar2007} or cell expansion/contraction in isogonal modes \cite{Noll2017}, which occur without requiring mechanical work. 
Besides this configurational degeneracy, it was shown in Ref.~\cite{Yang2017}, while studying the SVP model, that the homogeneous model~\eqref{eq-vertexEnergyHomogeneous} also presents a parameter degeneracy: the system dynamics does not depend on the precise value of $A_0$. This degeneracy is also present in  VM~\cite{Arzash2025}. The usual approach to addressing this issue is to fix $A_{0}$ to a specific value, such as, for example, to the average available space for cells, and work only with the remaining parameters. 

The homogeneous VMs are popular for their simplicity~\cite{Farhadifar2007, Hoevar2009, Staple2010,Mao2013, Bi2014, Bi2015, Park2015,Barton2017,Nestor-Bergmann2018, Tetley2019, Armon2021, Damavandi2025}.
Nevertheless, assuming \textit{mechanical homogeneity} for biological tissues is a strong claim. 
Given the inherent complexity of biology, there is no reason to believe that the diversity of cell shapes, sizes, and compositions would not also be accompanied by differences in cell stiffness and other mechanical properties~\cite{Mammoto2010, Mammoto2013}. Indeed, \cite{Parihar2025} reported pronounced mechanical heterogeneity across tissues derived from thirty cell lines and seven substrates, \cite{Nehls2019} observed a nonlinear increase in stiffness with cell area in epithelial cells, and the review  \cite{Moeendarbary2014} reaffirmed that cell morphology and composition are correlated with mechanical properties.

The mechanical heterogeneity in tissues can be accounted for in the VM by letting the stiffness and target parameters vary from cell to cell. Already in Ref.~\cite{Nagai2001}, this possibility was considered in the formulation of the VM, showing that model heterogeneity allows for a more realistic cell configuration than the homogeneous version. Also, heterogeneous formulations of the VM have been used in the description of epithelial morphology and morphogenesis~\cite{osterfield2013three,hannezo2014theory,misra2017complex,castle2025cell} and in the process of epithelial invagination~\cite{brezavvsvcek2012model,
misra2016shape}, where regions of the tissue have different mechanical properties. 
Cases where the tissue is fully heterogeneous have been analyzed in Refs.~\cite{Noll2017,Li2019,Kim2021} with heterogeneity in the perimeter and tension parameters but homogeneous area and pressure terms.
This leads to the question: in the heterogeneous case, is the target area also irrelevant? More precisely, can the parameter degeneracy in the target area be properly extended to an averaged value in heterogeneous systems while maintaining complete physical generality?

Here, we investigate in detail the parameter degeneracy associated with the target area in the heterogeneous VM. 
A first exploration was performed in Ref.~\cite{Arzash2025}, where they investigated the possibility of a heterogeneous target shape index as a control parameter for rigidity transitions.  The authors concluded that the value of the average target area $\langle A_{0c} \rangle$ did not (statistically) change the dynamics of the system. 
Here we show that, rather, it is the value of the average product between cell stiffness and target area, i.e., $\langle K_{Ac}A_{0c} \rangle$, that is irrelevant for the dynamics and that the result in Ref.~\cite{Arzash2025} is only approximate. This implies that, contrary to the case of the homogeneous model, where the target area parameter is entirely irrelevant for the dynamics, in heterogeneous models, the individual target areas do change the system dynamics, and by so, their parameter space is relevant. 

Important observables, such as the target shape index~\cite{Bi2015}, the cell pressure, and the Cauchy cell stress tensor \cite{Yang2017} depend, in their definition, on the target areas. 
Hence, the physical relevance of their numerical values depends on the target area degeneracy being properly treated or removed. We present a method to deal with the degeneracy, using the concept of gauge cell pressure, which allows to compare results obtained in studies performed with different set of parameters. 

The rest of the article is organized as follows. In Section \ref{sec-dyn}, we describe the heterogeneous VM, where the cell pressure is identified.  
In Section \ref{sec-PGT}, we present the methodology that allows for the identification of parameter space degeneracy and reveals the symmetry transformation underlying it. In Section \ref{sec-gaugePressure}, we show that this symmetry can be interpreted as a gauge cell pressure. Two different, but equivalent, methods are introduced to adequately fix the gauge and eliminate the degeneracy, allowing for comparison of the results using different sets of parameters. We also discuss how to deal with the degeneracy when the model parameters are optimized to reproduce experimental results. In Section \ref{sec-BCcurvature}, we examine how boundary conditions and tissue curvature may affect the parametric degeneracy. Finally, conclusions are presented in Section~\ref{sec.conclusions}.

\section{Vertex model} \label{sec-dyn}

\begin{figure}[htb]
    \centering
    \includegraphics[trim=1cm 7cm 1cm 1.5cm, clip, width=1.0\linewidth]{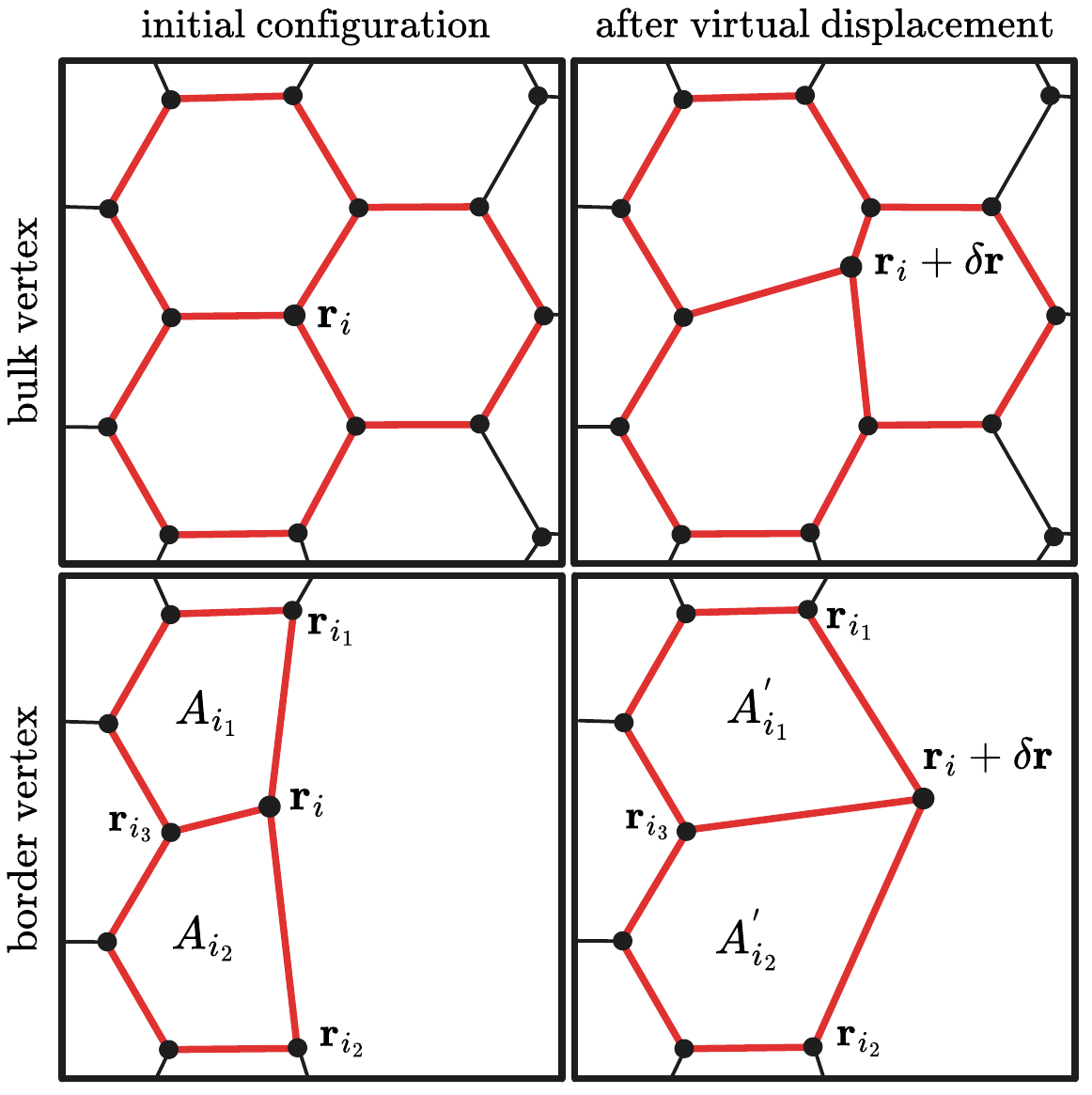}
    \caption{Tissue schematics showing the position $\mathbf{r}_i$ of a bulk vertex (top) and a border vertex (bottom), in their initial configurations (left column) and after a virtual displacement $\delta \mathbf{r}$ (right column). The displacement of the bulk vertex does not alter the net area of the cells it belongs to (highlighted in red), whereas the displacement of a border vertex does.}
    \label{fig-areaDegeProve}
\end{figure}

In the two-dimensional vertex model, cells are modeled as polygons, which cover the full domain with no overlap between them, describing confluent tissues. In this geometry, edges are limited by two vertices. In the bulk, cell edges are shared by two cells, and vertices by three or more cells and edges (see Fig.~\ref{fig-areaDegeProve}-top). For an open tissue, the border vertices are shared by only two cells (see Fig.~\ref{fig-areaDegeProve}-bottom). Also, no vertices are allowed inside an edge. 
Unless explicitly stated otherwise, we consider two-dimensional planar tissues composed of $N_v$ identical vertices, $N_c$ cells, and $N_e$ edges, in a rectangular domain with periodic boundary conditions. The extension to other geometries is treated in Sect.~\ref{sec-BCcurvature}. 
The dynamics of the tissue is given by the motion of the vertices, which follow overdamped variational equation of motion. For the vertex $i$,
\begin{equation}\label{eq-forceOverdamped}
    \frac{d \mathbf{r}_i}{d t} = \frac{1}{\gamma}\mathbf{F}_i = -\frac{1}{\gamma}\frac{\partial E}{\partial \mathbf{r}_i},
\end{equation}
where $\mathbf{r}_i$ is the vertex position, $\mathbf{F}_i$ is the total force acting on vertex $i$ obtained from the tissue energy functional $E$, and $\gamma$ is the homogeneous and constant viscosity coefficient.

Here we will work with an extension of the homogeneous model \eqref{eq-vertexEnergyHomogeneous}, to consider its heterogeneous  version, where each cell has independent parameters, 
\begin{equation}\label{eq-vertexEnergyHeterogeneous_2}
    E = \sum_c \left [ \frac{K_{Ac}}{2}\left (A_c - A_{0c} \right )^2 + \frac{K_{Lc}}{2}(L_c - L_{0c})^2 \right]. 
\end{equation}
The dynamics is such that the vertices move to make the cells approach their target areas $A_{0c}$ and target perimeters $L_{0c}$, which defines the system energy minimum, and the evolution towards this minimum happens with intensities $K_{Ac}$ and $K_{Lc}$, which give the system time scales.

In the vertex model, by computing the virtual work with a Hamiltonian interpretation of $E$, it is possible to obtain the  2D cell pressure $P_c$ of cell $c$ and the tension $T_k$ associated to each edge $k$, of length $l_k$ \cite{Chiou2012}:
\begin{align}
    \label{eq-E_derivative_total_2}
    dE\left( \left\{ A_c\right\},\left\{ l_k\right\}\right)
    =-\sum_c P_c dA_c + \sum_k T_k dl_k,
\end{align}
where the second sum is over all the edges in the tissue and
\begin{align}
    P_c &\equiv -\frac{\partial E}{\partial A_c} = - K_{Ac} (A_c - A_{0c}),  \label{eq-pressure-definition}\\
    T_{k} &\equiv \frac{\partial E}{\partial l_k} = \sum_{c / k\in c} \frac{\partial E}{\partial L_{c}}= \sum_{c / k\in c}K_{Lc} \left ( L_{c} - L_{0c} \right ), \label{eq-tensio-definition_2}
\end{align}
with the sum in Eq.~\eqref{eq-tensio-definition_2}  over the two neighbor cells whose shared membrane form the edge $l_k$.

\section{parameter space degeneracy}\label{sec-PGT}

To identify the degeneracy, we perform a general transformation in parameter space  to all $N_c$ cells of a heterogeneous  tissue:
\begin{align}\label{eq-PGT}
   A_{0c}^{'} &= A_{0c} + q_{Ac}, & K_{Ac}^{'} &= K_{Ac} + q_{K_{Ac}},\notag\\ 
   L_{0c}^{'} &= L_{0c} + q_{Lc}, & K_{Lc}^{'} &= K_{Lc} + q_{K_{Lc}}, 
\end{align}
where $\{q_{Ac},\,q_{K_{Ac}},\,q_{Lc},\,q_{K_{Lc}}\}$ are a total of $4N_c$ real numbers. We then calculate the force difference between the transformed system and the original one:
\begin{multline}
\label{eq-force_Diference}
    \mathbf{F}_i^{'} - \mathbf{F}_i = -\frac{\partial}{\partial \mathbf{r}_i} \left ( E^{'}-E \right ) \\
    = \frac{\partial}{\partial \mathbf{r}_i} \sum_{c / i\in c} K_{Ac} q_{Ac}  A_{c}
    -\sum_{c / i\in c}q_{K_{Ac}}\left ( A_{c} - {A_{0c}^{'}} \right )\frac{\partial}{\partial \mathbf{r}_i} A_{c} \\
    + \frac{\partial}{\partial \mathbf{r}_i} \sum_{c / i\in c} K_{Lc} q_{Lc} L_{c}
    -\sum_{c / i\in c}q_{K_{Lc}}\left ( L_{c} - {L_{0c}^{'}} \right )\frac{\partial}{\partial \mathbf{r}_i} L_{c},
\end{multline}
where the sums are over all the cells to which the vertex $i$ belongs. We look  for possible values $q_{Ac}^*,\,q_{K_{Ac}}^*,\,q_{Lc}^*,\,q_{K_{Lc}}^*$ that produce the same force, that is, are solutions for $\mathbf{F}_i^{'} - \mathbf{F}_i=0$, indicating that there is a degeneracy in parameter space. In models with additional non-variational forces, like activity \cite{Verdugo2022}, or models with tunable degrees of freedom \cite{Arzash2025}, one should modify Eq.~\eqref{eq-force_Diference} to include these extra terms.
The first term on the right side of Eq.~\eqref{eq-force_Diference} is zero when $q_{A_c}=q_{A_c}^*=P_0 / K_{Ac}$, for any $P_0$. Indeed, this term simplifies to $P_0\frac{\partial}{\partial \mathbf{r}_i} \sum_{c / i\in c} A_{c_i}$, but as shown in Fig.~\ref{fig-areaDegeProve}-top, variations in  $\mathbf{r}_i$ do not change the total area of the  participating cells, resulting in a null contribution from this term.  For the other parameters, $q_{K_{A_c}}$, $q_{L_c}$, and $q_{K_{L_c}}$,  they generally produce changes in the forces, except for the highly particular case of a regular tissue with all cells of equal area, in which case  for any value of $q_{K_{A_c}}$ the force does not change. In consequence, under general conditions, the heterogeneous vertex model is degenerated under the symmetry transformation
\begin{equation}\label{eq-vm_simetryHeterogeneous}
    A_{0c} \rightarrow A_{0c}^{'} = A_{0c} + \frac{P_0}{K_{Ac}},
\end{equation}
with the remaining parameters fixed. 
Note that the target area transformation must be applied to  all $N_c$ cells simultaneously, where $P_0$ can take any value, positive or negative. 
Although Fig.~\ref{fig-areaDegeProve}-top shows cases where three cell meet on a vertex, the derivation above does not depend on the number of cell at a vertex and the  symmetry~\eqref{eq-vm_simetryHeterogeneous} holds true for any type of bulk vertices, including four fold and rosette ones.
Finally, for a homogeneous tissue, the symmetry transformation reduces to $A_{0}\rightarrow A_{0}^{'} = A_{0} + a_0$,  
with $a_0=P_0/K_A$, which is equivalent to the result reported by~\cite{Yang2017} and is the reason why, for example, the rigidity transition point does not depend on the value of $A_0$ in the homogeneous model~\cite{Yang2017}.

%%%%%%%%%%%%%%%%%%%%%%%%
\section{Managing the  degeneracy}\label{sec-gaugePressure}
\subsection{Gauge pressure}\label{sec-gaugePressure_and_obsDege}

In Ref.~\cite{Yang2017}, it is shown that in a homogeneous model, changing the target area value only changes the offset of the tissue pressure. Similarly, here, we show that $P_0$ sets the global internal pressure of the cells in the tissue. 
Indeed, applying the transformation \eqref{eq-vm_simetryHeterogeneous} to Eq.~\eqref{eq-pressure-definition} results in that the cell pressure changes as 
\begin{equation}\label{eq-presueNotMatter_2}
    P^{'}_c = - K_{Ac} \left ( A_c - A_{0c} - \frac{P_0}{K_{Ac}} \right )
    = P_c + P_0.
\end{equation}
The average cell pressure $\langle P_c \rangle \equiv 1/N_c\sum_c P_c=-\langle K_{Ac}(A_c-A_{0c})\rangle$,
changes precisely by $P_0$. Then, it is possible to interpret that the symmetry transformation \eqref{eq-vm_simetryHeterogeneous} fixes the gauge for the global pressure. As a result, the cell pressure \eqref{eq-pressure-definition} is gauge dependent and should not be considered as a relevant physical observable. On the contrary, the relative cell pressure $\Delta P_c\equiv P_c - \langle P_c \rangle$ is invariant under the symmetry transformation. 
Also, the usual expression for the cell stress tensor considers the isotropic contribution of the cell pressure~\cite{Yang2017}
\begin{equation}\label{eq-stressTensor_lit}
\boldsymbol{\sigma}_c = - P_c \mathbf{ \mathbb{I} } + \frac{1}{2 A_c} \sum_{k \in c} T_k\frac{\mathbf{l}_k \otimes \mathbf{l}_k}{l_k},
\end{equation}
and, therefore, the stress tensor is also gauge dependent. Here, $\mathbb{I}$ is the identity matrix, the sum is over all the edges $k$ of cell $c$, and $\mathbf{l}_k=l_k\mathbf{\hat{l}}_k$ is the edge vector. The solution comes from replacing $P_c$ with $\Delta P_c$ in its definition, which is equivalent to fixing the stress tensor gauge, as explained in the next section and the details will be published elsewhere.

The average cell pressure can be written as
\begin{equation}\label{eq-aveCellPressure}
    \langle P_c \rangle = -\left < K_{Ac} A_c \right > + P_g, 
\end{equation}
where the first term on the right hand side depends on the dynamical degrees of freedom, while 
\begin{equation}\label{eq.gaugepressure}
P_g  \equiv  \langle K_{Ac} A_{0c} \rangle 
\end{equation} 
depends only on model parameters that we call \textit{gauge pressure}, since $P_g$ is the source of pressure degeneracy [under the transformation \eqref{eq-vm_simetryHeterogeneous}, $P_g^{'}=P_g+P_0$]. 
For homogeneous models, the gauge pressure simplifies to $P_g=K_AA_0$, where $A_0$ is physically \textit{irrelevant} for the system dynamics.

For heterogeneous models, Eq.~\eqref{eq.gaugepressure} implies that the value of $\sum_c K_{Ac} A_{0c}$ is irrelevant for the dynamics, but the individual target area parameters do affect the evolution of the tissue.
Hence, while for the homogeneous case, the target area parameter space is totally irrelevant, in the heterogeneous model, the target area is only irrelevant in the parameter space regions defined by Eq.~\eqref{eq-vm_simetryHeterogeneous}. For example, in the heterogeneous case, changing all target areas by the same constant value will change the dynamics. 

A reliable comparison of model results requires the systems to be studied under the same gauge. 
Using the transformation~\eqref{eq-vm_simetryHeterogeneous}, $P_g$ can be set to \textit{any} value without affecting the system dynamics. 
This symmetry enables comparisons between systems originally analyzed with different gauges. 
In particular, a system with gauge pressure $P_g$ can be transformed to a new gauge $P_g'$ through

\begin{align}\label{eq-guagePressure_transformation}
    A_{0c} \rightarrow A_{0c}^{'} = A_{0c} + \frac{P_g^{'}\,-\,P_g}{K_{Ac}},
\end{align}
where $P_g^{'}$ can be chosen as any value one may find convenient.

As some observables (the cell pressure or the target shape index, for example), depend in their definitions on the target areas, to obtain a fair comparison between different systems, these systems should be put in the same gauge by the transformation \eqref{eq-guagePressure_transformation}. This gauge specification can be done before running a simulation for example or, as the symmetry transformation does not change the dynamics, can be done at the end.

\subsection{Fixing the gauge pressure}\label{subsec-ZPG}
For the study of homogeneous models, as the target area is irrelevant, it has become usual to fix it to some specific value, following some criteria that vary among papers~\cite{Nagai2001,Barton2017,Tetley2019,Armon2021,Tah2025}. 
In rigidity-theory-oriented works and other articles, it has become usual to fix it to the average available area per cell in the system $A_0=A_\text{total}/N_c$~\cite{Nagai2001,Bi2016, Yang2017,Barton2017,Sussman2018,Sussman2018-PRL,Li2019,Wang2020}. 
Also, in these studies time units are normally fixed by setting $K_A=1$ and length units with $A_\text{total}/N_c=1$, implying that the gauge pressure is $P_g=1$ in these units.
In this gauge, the average pressure $\langle P_c \rangle=K_A(\langle A_c\rangle -A_0)$  vanishes identically in confluent tissues with fixed total area because, under this condition, $\langle A_c\rangle=A_\text{total}/N_c$.  Here, we investigate how this zero-averaged pressure gauge (ZPG) reads for the case of heterogeneous models. 
Eq.~\eqref{eq-aveCellPressure} implies that to get $\langle P_c \rangle=0$, the gauge pressure must be set to $P_g^{\text{ZPG}}=\left < K_{Ac} A_c \right >$, which will generally depend on time. Using Eq.~\eqref{eq-guagePressure_transformation}, this means the target areas should be time-dependent as well. 
In a simulation, if the initial distribution of target areas are given by $A_{0c}^\text{ini}$, the target areas should evolve on time as
\begin{align}\label{eq-timeDependetTargetArea-1}
    A_{0c}^{\text{ZPG}}(t) = A_{0c}^\text{ini} + \frac{\left < K_{Ac} A_c(t) \right > -\left < K_{Ac} A_{0c}^\text{ini} \right >}{K_{Ac}},
\end{align}
which directly give $\langle P_c \rangle=0$. Also, as the symmetry changes do not change the dynamics, it is also possible to run a simulation with a fixed distribution of target areas and, at the end, transform all parameters with Eq.~\eqref{eq-timeDependetTargetArea-1} to compute the relevant observables. 
In this gauge, $P_c^{\text{ZPG}}=\Delta P_c=- K_{Ac} (A_c - A_{0c}^{\text{ZPG}})$, i.e., the pressure is automatically invariant, and we can see that replacing $P_c$ by the $\Delta P_c$ in Eq.~\eqref{eq-stressTensor_lit} is equivalent to fixing the stress tensor at the ZPG.

The election to fix $A_0=A_\text{total}/N_c$ in homogeneous models can be interpreted also as fixing the gauge pressure to $P_g=K_A A_\text{total}/N_c$. Extending this choice of setting $P_g$ to a fixed value to the case of heterogeneous models, implies that  the target areas are given by
\begin{equation}\label{eq-targetAreaFGP-1}
    A_{0c} = A_{0c}^\text{ini} + \frac{P_g^\text{fix}-\langle K_{Ac} A_{0c}^\text{ini} \rangle }{K_{Ac}},
\end{equation}
which are now time-independent. In this gauge, however, the average pressure  $\langle P_c \rangle(t) = -\left < K_{Ac} A_c(t) \right > + P_g^\text{fix}$ is not constant, and the relative cell pressure is given by $\Delta P_c=-K_{Ac}(A_c -A_{0c}) + \langle K_{Ac}(A_c -A_{0c})\rangle$. A simple possible choice is $P_g^\text{fix}=0$, but other elections are possible, for example $P_g^\text{fix} = \langle K_{Ac}\rangle A_\text{total}/N_c$ to be close to the usual gauge in homogeneous models.

For the homogeneous model, the control parameter of the rigidity transitions (called here target shape index) is usually defined as~\cite{Bi2015}
\begin{equation}\label{eq-shapeindex}
    p_0 = \frac{L_0}{\sqrt{A_0}}.
\end{equation}
In the ZPG, this reads as $p_0={L_0}/{\sqrt{A_\text{total}/N_c}}$ and it has been shown that for this gauge, $p_0^*=3.81$ marks the rigidity transition~\cite{Bi2015,Bi2016}. However, this numerical value depends on the selected gauge. Indeed, if we arbitrarily  chose for example $A_0=A_\text{total}/(4N_c)$, the transition point will still happen for $L_0=3.81\sqrt{A_\text{total}/N_c}$, implying that the critical target shape index~\eqref{eq-shapeindex}, will now be $p_0^*=3.81\sqrt{4}=7.62$. 
Note that in parallel with the study of the target shape index, simulations and experiments show that the average of the dimensionless cell perimeter $\bar{q}=\langle L_c/\sqrt{A_c}\rangle$, called here as geometric shape index, also marks the rigidity transition at $\bar{q}^*=3.81$ and therefore acts as an effective control parameter~\cite{Park2015,Bi2016,Yang2017}. The names used for $p_0$ and $\bar{q}$ vary among papers, with no consensus in the literature, but a critical difference is that  $p_0$ depends on the pressure gauge, while $\bar{q}$ is gauge-independent. 
In exploring potential control parameters analogous to the target shape index for heterogeneous models, as in Ref.~\cite{Arzash2025}, special care must be taken in their definition to ensure gauge-independent expressions.

In Ref.~\cite{Arzash2025}, while studying a heterogeneous VM with fixed density and periodic boundary conditions, the authors fixed the value of $\langle A_{0c} \rangle$, arguing that for homogeneous $K_A$, shifting $\langle A_{0c} \rangle$ by a constant $\Delta \langle A_{0c} \rangle$ introduced dynamical contributions that were negligible in simulations compared to those arising from other terms in the elastic energy. Indeed, if there is a unique $K_A$ for all cells, $\langle A_{0c} \rangle$ fixes the gauge pressure and changing its value does not introduce any dynamical contribution to the system.
Nevertheless, in \cite{Arzash2025}, they explored scenarios in which $K_{Ac}$ was not Dirac delta distributed, but rather $K_{Ac}=\langle K_{Ac} \rangle+\delta K_{Ac}$, with $\delta K_{Ac}$ coming form a distribution with a finite width $\sigma_{K_A}$. Changing all $A_{0c}$ by the same constant value $\Delta \langle A_{0c} \rangle$ gives for the force difference
\begin{align}    
    \mathbf{F}^{'}_i - \mathbf{F}_i =& \Delta \langle A_{0c}\rangle \frac{\partial}{\partial \mathbf{r}_i} \sum_{c/ i\in c} K_{A_c} A_c \notag \\ 
    =& \Delta \langle A_{0c}\rangle \langle K_{Ac} \rangle {\frac{\partial}{\partial \mathbf{r}_i} \sum_{c/ i\in c} A_c} \notag \\
    &+ \Delta \langle A_{0c}\rangle \sum_{c/ i\in c} \delta K_{A_c} \frac{\partial A_c}{\partial \mathbf{r}_i}.
\end{align}
The first term  in the second line cancels as usual. To estimate the  second term we use that $\delta K_{A_c} = \mathcal{O}(\sigma_{K_A})$, and $\frac{\partial A_c}{\partial \mathbf{r}_i} = \mathcal{O}(l)$, with $l$ the typical edge length (see Appendix~\ref{sec-apendix-force-equations}). Therefore, $ \mathbf{F}^{'}_i - \mathbf{F}_i \sim \Delta \langle A_c \rangle  \sigma_{K_A}  l$.
On the other hand, from Eq.~\eqref{eq-areaDerivada}, $\mathbf{F}_i = \mathcal{O}(\langle K_{A_c} \rangle   \langle A_c \rangle   l)$, yielding the relative force difference
\begin{equation}
    \frac{\mathbf{F}^{'}_i - \mathbf{F}_i}{\mathbf{F}_i} \sim  \frac{\Delta \langle A_{0c} \rangle}{\langle A_c \rangle}  \frac{\sigma_{K_A}}{\langle K_{A_c} \rangle},
\end{equation}
which, in general, is not null and in fact the only quantity that can be fixed to arbitrary values without changing the dynamics is the gauge pressure \eqref{eq.gaugepressure}.

\subsection{Parameter optimization procedures} \label{sec.optmization}

Whether fitting model parameters to reproduce experimental data or analyzing theoretical models to identify different system phases (e.g., solid or liquid), parameter space exploration is always a key component. The methods used for exploration vary — including Monte Carlo minimization, Monte Carlo Markov Chain, gradient descent, and grid-based approaches — but their essence is the same: to change one or several model parameters by a given amount and evaluate the resulting changes in the system's behavior. In the case of the VM, the transformation \eqref{eq-vm_simetryHeterogeneous} does not change the outcome but results in different model parameters. This can give discrepant results, which may be misinterpreted. Also, changing parameter values along the symmetry \eqref{eq-vm_simetryHeterogeneous} are computationally inefficient since they do not change system dynamics.

In the homogeneous case,  eliminating parameter motion along the symmetry is simply achieved by fixing  $A_0$. In the heterogeneous case, every move that changes $K_{Ac}$ or $A_{0c}$ for one or several cells modifies the system's gauge pressure. Thus, in this case one must reapply Eqs.~\eqref{eq-timeDependetTargetArea-1} or \eqref{eq-targetAreaFGP-1} to rescale $A_{0c}$ accordingly, maintaining the desired gauge. This construction allows one to work with arbitrary distributions of $A_{0c}$ and $K_{Ac}$, as long as the results of different systems are compared within the same gauge.

Inference methods can be used to fit the model parameters of homogeneous vertex models to experimental data (see, for example~\cite{farhadifar2007influence,mao2011planar,spahn2013vertex,Mao2013,henkes2020dense,Ogita2022,Verdugo2022}). For heterogeneous models, if sufficient experimental data is provided, it could be possible in principle to obtain all target areas $A_{0c}$ and stiffnesses $K_{Ac}$, up to the combination giving the gauge pressure $P_g=\langle K_{Ac}A_{0c}\rangle$, which must be fixed a prior. 

\section{Boundary conditions and Curvature}\label{sec-BCcurvature}

\subsection{Finite tissues}
Real tissues are not periodic or infinitely large. Here we discuss how the degeneracy manifests in the case of three different boundary conditions. 
First, we consider border vertices that are fixed or have an imposed but not necessarily constant velocity $\mathbf{v}_B(\mathbf{r},t)$ \cite{Reig2017, Verdugo2022}. This last scenario can appear when modeling an experimental system in which the border dynamics is unknown because other forces are in action (as in the margin of the enveloping layer during epiboly~\cite{Reig2017}) or the experimental field of view is limited. 
Then, by providing the velocity at the border with experimental data, it is possible to simulate and describe the dynamics in the bulk of the tissue. 
In this case, the bulk vertices are described with the force field of the VM  but the border vertices are simply described by $\mathbf{F}_{B}=\gamma \mathbf{v}_B(\mathbf{r},t)$ ($\mathbf{v}_B=0$ for fixed borders).  The transformation of the target areas by Eq.~\eqref{eq-vm_simetryHeterogeneous} will therefore not change the forces acting on any vertex and the parameter degeneracy is maintained. 

Second, in a tissue with free boundaries $B$ (open tissue), the force difference in Eq.~\eqref{eq-force_Diference} when changing the target areas is
\begin{equation}\label{eq-ForceDifferenceOpenTissue}
    \mathbf{F}_i^{'} - \mathbf{F}_i 
    \sim 
    P_0 \frac{\partial}{\partial \mathbf{r}_i} \sum_{c / i\in c} A_{c}
    =
    \begin{cases}
    0, & i \notin B \\ 
    P_0 \frac{\partial}{\partial \mathbf{r}_i}\left [  A_{i_1} + A_{i_2} \right ] , & i \in B 
    \end{cases},
\end{equation}
where $A_{i_1}$ and $A_{i_2}$ are the areas of the two cells shared by the boundary vertex $i$ (see Fig.~\ref{fig-areaDegeProve}-bottom).
Although the degeneracy is maintained for all the bulk vertices ($i\notin B$), it is removed for all border vertices as shown schematically in the figure. Hence, the transformation of the target areas by Eq.~\eqref{eq-vm_simetryHeterogeneous} will change the dynamics of the system and will manifest as a pressure gradient generated at the border and propagating through the interior of the tissue.  
In this case, if the model parameters are optimized as discussed in Sect.~\ref{sec.optmization}, this \textit{partial degeneracy} manifests in that the error function used in an optimization process will presents a shallow minimum in parameter space.

Third, if there is an additional external tension acting on the border edges $T_{k}^{\text{ext}}$, an extra term should be added to $E$
\begin{equation}\label{eq-vertexEnergyHomogeneousTnesionExtra}
    E_B = \sum_{k\in B}T_{k}^{\text{ext}}l_{k}.
\end{equation}
By making the transformation $T_k^{\text{ext}}\rightarrow T_k^{'{\text{ext}}}=T_k^{\text{ext}}+\delta T_k^{\text{ext}}$ and the target area transformation \eqref{eq-vm_simetryHeterogeneous}, gives for the force difference of the vertices at the border
\begin{equation}\label{eq-ForceDifferenceOpenTissue_tensionBorder}
    \left( \mathbf{F}_i^{'} - \mathbf{F}_i \right )_{i \in B} 
    =
    \frac{P_0}{2} \left ( \mathbf{r}_{i_1} - \mathbf{r}_{i_2} \right ) \times \mathbf{\hat{z}} - \delta T_{i_1} \mathbf{\hat{r}}_{i,i_1} - \delta T_{i_2} \mathbf{\hat{r}}_{i,i_2},
\end{equation}
where $i_1$ and $i_2$ label the two neighbor vertices of $i$ that are also $\in B$ (see Fig.~\ref{fig-areaDegeProve}-bottom), and $\mathbf{\hat{r}}_{i,i_{1,2}}$ are unit vectors pointing along $\mathbf{r}_{i_{1,2}}-\mathbf{r}_i$. Then, the target area degeneracy could be maintained if the edge tensions change also such that Eq.~\eqref{eq-ForceDifferenceOpenTissue_tensionBorder} vanishes for all vertices in the border. However, this is generally not possible to satisfy. Indeed, for a closed boundary with $N_b$ vertices and edges there are $N_b$ values of $\delta T_k^{\text{ext}}$ to fix, but for each vertex there are two independent equations (the $x$ and $y$ components), resulting in $2(N_b-1)$ equations (there is one redundant equation), and the system of equations is overdetermined. Consequently, there is no general transformation that keeps the forces unchanged and the degeneracy is removed. As in the case of open boundaries, the force difference in the bulk vertices vanishes and this \textit{partial degeneracy} will give rise to a shallow minimum in parameter space for the error function if the model parameters are optimized.

\subsection{Curved tissues}\label{sec-degener-curvedspaces}
During both animal development and homeostasis, tissues rarely organize into simple flat sheets; instead, they form complex, curved three-dimensional structures. For instance, adult animal eyes are typically spherical, blood vessels are cylindrical, brains exhibit folds with varying curvature, or embryos progress from a spherical blastula to an invaginated gastrula.
A particularly relevant case is that of the early development of animal embryos where, in the epiboly, a quasi planar enveloping layer moves on top of an almost spherical yolk. 
VMs have been used to describe tissues in curved geometries — for example, during epiboly~\cite{Weliky1990,Reig2017,Verdugo2022}, neural tube formation~\cite{Odell1981}, ventral furrow formation during gastrulation~\cite{brezavvsvcek2012model}, epithelial folding in the imaginal leg disc~\cite{Monier2015}, and more~\cite{Alt2017}. 
In such curved spaces, it is also necessary to account for the target area degeneracy. 

Although Eq.~\eqref{eq-force_Diference} does depend on the metric of the space (via $A_c$ and $L_c$), in confluent tissues, $\frac{\partial}{\partial \mathbf{r}_i} \sum_{c / i\in c} A_{c_i}$  still vanishes since the sum of the areas does not depend on the position of the shared vertex. Hence, the degeneracy remains valid for curved tissues as well.
However, under certain circumstances, the cell areas $A_c$ and perimeters $L_c$ are not computed exactly, and numerical approximations are used in simulations to calculate these values. In these conditions, the sum of the areas of cells sharing a vertex can depend on the vertex position, and the degeneracy is 
artificially suppressed. Let us focus on a tissue over a sphere. The calculation of $L_c$ and $A_c$ requires the use of geodesic paths and spherical triangles, which can be cumbersome to work with. An alternative is to use \textit{local planar approximations}~\cite{osterfield2013three,Verdugo2022},
where the edges $l_k$ are defined as the Euclidean distance between the two vertices of the edge $k$, and $A_c$ is calculated by replacing $\mathbf{\hat{z}}$ in Eq.~\eqref{eq-areaDerivada} by the cell center unit vector $\widehat{\mathbf{R}}_c=\frac{\mathbf{R}_c}{|\mathbf{R}_c|}$, which is the normal vector of the plane associated with cell $c$. 
Now, when applying the target area transformation \eqref{eq-vm_simetryHeterogeneous}, the force difference for a spherical tissue with local planar approximation reads
\begin{align}
    \mathbf{F}_i^{'} - \mathbf{F}_i 
    &    = P_0 \sum_{c / i\in c} \frac{\partial A_{c}}{\partial \mathbf{r}_i} \notag\\
    &= -\frac{P_0}{2} \sum_{c / i\in c}\widehat{\mathbf{R}}_{c}\times\left ( \mathbf{r}_{i_{c}+1}-\mathbf{r}_{i_{c}-1} \right ) \notag\\
    & \sim 3 \frac{P_0}{2} \frac{D}{R},
\end{align}
where in going to the second line we replaced $\frac{\partial A_{c}}{\partial \mathbf{r}_i}$ using Eq.~\eqref{eq-areaDerivada}. For the final estimate, $D$ is the average cell diameter and $ R$ is the sphere radius. For large radii, the force difference is small, and again, if the model parameters are optimized against some experimental data, this would lead to a shallow minimum in an error function.

Even for small values of $D/R$, the partially degenerated system described above can be troublesome since it hinders the possibility of applying Eqs.~\eqref{eq-timeDependetTargetArea-1} and \eqref{eq-targetAreaFGP-1} to treat the degeneracy correctly. In some cases, it can lead to convergence failure when inferring parameters. Fortunately, a heuristic \textit{quasi-spherical} approach can be used to restore the system to complete degeneracy
%.
The force on a vertex $i$ can be also expresses as the sum $\mathbf{F}_i=\sum_{j\sim i}\mathbf{F}_{ij}$ of the forces due to all edges that join the vertex $i$ with vertices $j$~\cite{Chiou2012}.
For a planar two-dimensional tissue (in the $XY$ plane), the pressure and tension components of the forces read as ($\mathbf{F}_{ij}=\mathbf{F}_{ij}^P + \mathbf{F}_{ij}^T$)
\begin{align}   
    \mathbf{F}_{ij}^P&=\frac{1}{2} \left ( P_{ij}^{l} - P_{ij}^{r} \right )\,r_{ij}\,\mathbf{\hat{n}}_{ij}, \label{eq-Fij}\\
     \mathbf{F}_{ij}^T&=T_{ij}\,\mathbf{\hat{r}}_{ij},   
\end{align}
where $\mathbf{\hat{n}}_{ij}\equiv \mathbf{\hat{r}}_{ij}\times\mathbf{\hat{z}}$ is the normal vector to the edge surface, $\mathbf{\hat{r}}_{ij}$ is the edge unity vector pointing from vertex $i$ to vertex $j$, and $\mathbf{\hat{z}}$ is the unit normal vector to the $XY$ plane. Here, $P_{ij}^l$ and $P_{ij}^r$ are the pressures of the cells given by Eq.~\eqref{eq-pressure-definition} on the left side and the right sides of the edge, defined according to the orientation of $\mathbf{\hat{r}}_{ij}$, and $T_{ij}$ is the tension of edge $ij$. It is direct that $\mathbf{F}_{ij}^P$ is free of degeneracy. 
On the surface of a sphere, Eq.~\eqref{eq-Fij} can be generalized to 
\begin{equation}
\mathbf{F}_{ij}^{P} = \frac{1}{2}\left ( P_{ij}^{l} - P_{ij}^{r} \right ) \, l_{ij} \, \mathbf{\hat{t}}(\theta_i, \phi_i)\times \mathbf{\hat{r}}(\theta_i, \phi_i), \label{eq.FP-sphere}
\end{equation}
where $l_{ij}$ is the geodesic length of edge $ij$, $\mathbf{\hat{t}}$ is the unit tangent vector along the geodesic edge $ij$, and $\mathbf{\hat{r}}$ is the unit radial vector, both evaluated at the vertex position. A key advantage of this formulation is that the resulting force now depends explicitly on the vertex position vectors $\mathbf{\hat{r}}_i$, rather than on cell centroids as in the planar approximation. As a consequence, the force equations retain the degeneracy and thus Eqs.~\eqref{eq-timeDependetTargetArea-1} and \eqref{eq-targetAreaFGP-1} can be applied to treat the degeneracy.
Using Eq.~\eqref{eq.FP-sphere} to compute the forces, restores the target area degeneracy regardless if the cell areas are computed using a local planar approximation. Additional precision can be gained computing $A_c$ as a spherical polygon using Girard's theorem~\cite{todhunter1886}.

\section{Conclusions and discussion}\label{sec.conclusions}

The vertex model has gained increasing attention in the study of biological tissues thanks to its simplicity and ability to describe several phenomena, for example, in animal development.
In its simplest, homogeneous formulation, the tissue is characterized by equal stiffness and target parameters for all cells. However, biological tissues are normally heterogeneous with cells differing in their mechanical properties. Here, we investigate how the reported irrelevance of the target area parameter in homogeneous models reads in the case when the tissue is heterogeneous. 

We found that in heterogeneous models, the target areas and stiffnesses of individual cells are relevant parameters that determine the system dynamics, but their weighted average $\langle K_{Ac}A_{0c}\rangle$ is irrelevant and its value can be changed arbitrarily without modifying the dynamics. This average, however, has the effect of changing the average cell pressure of the system and, therefore, can be identified as setting the gauge pressure. We show then that the relative cell pressure, defined as the difference of the cell pressure to the average, is a well defined observable that is invariant to the found parameter degeneracy.
Alternatively, if it is opted to use an  observable that depends on the pressure gauge, a proper analysis and comparison between models and systems must be done in a common gauge. For that, we provide methods to fix the gauge even in cases where the model parameter are being optimized, for example, to reproduce experimental results. 

Besides the pressure, the cell stress tensor and the target shape index also depend on the target area, and, therefore, their values depend on the chosen pressure gauge. We identify that the usual gauge election in the analysis of rigidity transitions corresponds to setting the average cell pressure to zero. Based on this, we propose possible elections to reproduce the zero-averaged pressure gauge in heterogeneous models.

For heterogeneous systems, the appropriate control parameter for the rigidity transition remains under debate~\cite{Arzash2025,Damavandi2025,Li2019}. Nevertheless, care should be taken to use gauge-independent definitions or, alternatively, explicitly indicate the chosen gauge. For example, in reference \cite{Arzash2025}, which analyzed confluent tissues with a fixed number of cells and total area, they proposed two candidates: $p_{0}^\text{I}=\langle L_{0c}/\sqrt{A_{0c}} \rangle$ and $p_{0}^\text{II}=\langle L_{0c}\rangle/\sqrt{\langle A_c\rangle}$. Moreover, since the authors used $\langle A_c\rangle=\langle A_{0c}\rangle$, a third option $p_{0}^\text{III}= \langle L_{0c}\rangle/\sqrt{\langle A_{0c} \rangle}=p_{0}^\text{II}$ is also implicit in the their analysis. Of those, $p_{0c}^\text{II}$ is gauge-independent, but is a \textit{mixed quantity} because it depends on parameters (via $L_{0c}$) and on geometrical properties (via $\langle A_{c}\rangle$).

We recall that the parameter degeneracy we describe, either in homogeneous or heterogeneous models, is not related to the freedom in choosing units. Indeed, the usual election in homogeneous systems, $K_A=1$ and $A_0=1$, to fix time and length units, which normalizes energy by $K_A A_0^2$~\cite{Staple2010,Bi2015}, still does not fix the gauge pressure, which is dimensional. The latter can only be properly fixed (in the homogeneous case) by setting \(A_0\) in comparison to other lengths in the system, e.g., the average available space per cell \(A_\text{total} / N_c\). In the heterogeneous case, the gauge can only be fixed via precise definitions of the cells' target areas, obtained through symmetry transformations in the parameter space, e.g., Eqs.~\eqref{eq-timeDependetTargetArea-1} and \eqref{eq-targetAreaFGP-1}.

Finally, the analysis presented here focused on the 2D vertex model with quadratic area and perimeter terms. Extensions to other models can be carried out in a similar way to the cases discussed for different boundary conditions and spherical domains. Future research directions include models that introduce additional tunable degrees of freedom~\cite{Arzash2025}, sub-rules of interaction at cellular edges~\cite{Noll2017}, curved membranes~\cite{Kim2021}, or activity and plasticity~\cite{Barton2017,Verdugo2022}, as well as investigating how the homogeneous self-propelled Voronoi model degeneracy~\cite{Yang2017} manifests in the heterogeneous case. Quadratic area terms with target areas are present in other cell–tissue models, such as the deformable particle model~\cite{Teixeira2021,Ourique2022,Teixeira2024}, raising the question of whether analogous degeneracies may arise there. 

\begin{acknowledgments}
This research was supported by the Fondecyt Grant No.\ 1220536 and Millennium Science Initiative Program NCN19\_170 of ANID, Chile. P.C.G. expresses his gratitude to the Chilean agency ANID for the grant No.\ 21231227, and the Brazilian agencies CAPES and FAPERGS for their financial support.
L.G.B. thanks CNPq for the grant 443517/2023-1. 
We thank Emanuel Teixeira, Silke Henks, and Carine Beatrici for inspiring discussions, Pablo de Castro and Lisa Manning for the encouragement, and Miguel Concha and the LEO group for discussions that inspired this work.
\end{acknowledgments}

%\bibliography{references}

\providecommand{\noopsort}[1]{}\providecommand{\singleletter}[1]{#1}%

\newpage

\appendix

\section{Force expressions}\label{sec-apendix-force-equations}

In a planar two dimensional tissue, the force on the vertex $i$ in Eq.~\eqref{eq-forceOverdamped} can be expressed as $\mathbf{F}_i=\sum_{c / i\in c}\mathbf{f}_i^c$, where the sum is over the three cells to which the vertex belongs and $\mathbf{f}_i^c$ is the force contribution coming from each cell,
\begin{equation}\label{eq-forceCellUnderived}
    \mathbf{f}_i^c = -K_{Ac}\left ( A_c - A_{0c} \right )\frac{\partial A_c}{\partial \mathbf{r}_i} - K_{Lc}\left ( L_c - L_{0c} \right )\frac{\partial L_c}{\partial \mathbf{r}_i},
\end{equation}
with
\begin{align}
    \frac{\partial A_c}{\partial \mathbf{r}_i}&=-\frac{1}{2}\mathbf{r}_{i_c-1,i_c+1}\times\hat{\mathbf{z}}, \label{eq-areaDerivada}\\
    \frac{\partial L_c}{\partial \mathbf{r}_i}&=-\hat{\mathbf{r}}_{i_c,i_c+1}+ \hat{\mathbf{r}}_{i_c,i_c-1}. \label{eq-perimDerivada}
\end{align}
Here, $\mathbf{r}_{i_c-1,i_c+1}$ is the vector pointing from the previous neighbor vertex of $i$ to its next neighbor, in cell $c$, and in clockwise order; $\hat{\mathbf{z}}$ is the unit normal vector perpendicular to the $x-y$ plane; and $\hat{\mathbf{r}}_{i,i\pm1}$ are the unit vectors pointing from vertex $i$ to its neighbors vertices $i\pm1$ in cell $c$ (see Fig.~\ref{fig-areaDegeProve}). 

\end{document}